\documentclass[prl,twocolumn,superscriptaddress,amsmath,amssymb,showpacs,floatfix,preprintnumbers]{revtex4-2}

\usepackage{amsmath}
\usepackage{graphicx}
\usepackage{lipsum}
\usepackage{dcolumn}
\usepackage{xcolor}
\usepackage{physics}
\usepackage{xr}
\usepackage{siunitx}
\usepackage{tabularx}
\usepackage{makecell}
\usepackage{multirow}
\DeclareGraphicsExtensions{.png .jpg .pdf}

\usepackage[normalem]{ulem}

\makeatletter
\newcommand*{\addFileDependency}[1]{% argument=file name and extension
  \typeout{(#1)}
  \@addtofilelist{#1}
  \IfFileExists{#1}{}{\typeout{No file #1.}}
}
\makeatother
 
\newcommand*{\myexternaldocument}[1]{%
    \externaldocument{#1}%
    \addFileDependency{#1.tex}%
    \addFileDependency{#1.aux}%
}

\myexternaldocument{si}

\begin{document}

\title{Metallic Electro-Optic Effect in Twisted Double-Bilayer Graphene}

\author{D. J. P. de Sousa}%\thanks{These authors contributed equally to this work.}
\email{sousa020@umn.edu}
\affiliation{Department of Electrical and Computer Engineering, University of Minnesota, Minneapolis, Minnesota 55455, USA}
\author{N. Roldan-Levchenko}\thanks{These authors contributed equally to this work.}
\affiliation{School of Physics and Astronomy, University of Minnesota, Minneapolis, Minnesota 55455, USA}
\author{C. O. Ascencio}\thanks{These authors contributed equally to this work.}
\affiliation{School of Physics and Astronomy, University of Minnesota, Minneapolis, Minnesota 55455, USA}
\author{J. D. S. Forte}
\affiliation{Department of Electrical and Computer Engineering, University of Minnesota, Minneapolis, Minnesota 55455, USA}
\author{Paul M. Haney}
\affiliation{Physical Measurement Laboratory, National Institute of Standards and Technology, Gaithersburg, Maryland 20899-6202, USA}

\author{Tony Low}
\affiliation{Department of Electrical and Computer Engineering, University of Minnesota, Minneapolis, Minnesota 55455, USA}
\affiliation{Department of Physics, University of Minnesota, Minneapolis, Minnesota 55455, USA}

\begin{abstract}
Recent theoretical advances have highlighted the role of Bloch state intrinsic properties in enabling unconventional electro-optic (EO) phenomena in bulk metals, offering novel strategies for dynamic optical control in quantum materials. Here, we identify an alternative EO mechanism in bulk metallic systems that arises from the interplay between Berry curvature and the orbital magnetic moment of Bloch electrons. Focusing on twisted double-bilayer graphene (TDBG), we show that the enhanced intrinsic properties of moir\'e Bloch bands give rise to a sizable linear magnetoelectric EO response, a first-order, electric-field-induced non-Hermitian correction to the gyrotropic magnetic susceptibility. This mechanism dominates in  $C_{3z}$-symmetric TDBG, where EO contributions originating from the Berry curvature dipole (BCD) are symmetry-forbidden. Our calculations reveal giant, gate-tunable linear and circular dichroism in the terahertz regime, establishing a robust and tunable platform for ultrafast EO modulation in two-dimensional materials beyond the BCD paradigm.

\end{abstract}
%\pacs{71.10.Pm, 73.22.-f, 73.63.-b}

\maketitle

%\textbf{key points} 
%- proposes a non-linear magnetoelectric effect
%- proposes a means to measure bias induced gyrotropic magnetic effect on the Fermi surface
%- identifies twisted double layer graphene as a system displaying giant non-linear 
%magnetoelectric responses. The effect is leading order in this system, provided C3z is there.
%- predicts light amplification through this effect, serving as a means to detect and for future applications. 

Noncentrosymmetric metals have recently emerged as a fertile platform for unconventional electro-optic (EO) phenomena, revealing rich low-frequency responses previously thought to be suppressed in bulk metallic systems~\cite{PhysRevLett.130.076901, Ma2025, PhysRevB.109.245126, PhysRevApplied.22.L041003, https://doi.org/10.48550/arxiv.2502.03399, PhysRevB.108.L201404}. In such materials, the intrinsic properties of Bloch states on the Fermi surface, such as the Berry curvature and the orbital magnetic moment, can couple static and optical fields in a fundamental manner, giving rise to non-reciprocal effects in the linear optical regime~\cite{PhysRevB.108.L201404, PhysRevB.109.245126}. While previous studies have primarily focused on EO effects mediated by the Berry curvature dipole (BCD)~\cite{PhysRevLett.130.076901}, new theoretical developments have revealed that the orbital magnetic moment texture of Bloch electrons on the Fermi surface can generate an entirely distinct class of EO responses, of which the so-called magnetoelectric EO effects is present in time-reversal symmetric systems~\cite{PhysRevB.108.L201404}. However, the effects remain largely unexplored in realistic material platforms, and their role in enabling optical control have yet to be demonstrated or quantified.

\begin{figure*}[t]
\includegraphics[width = \linewidth]{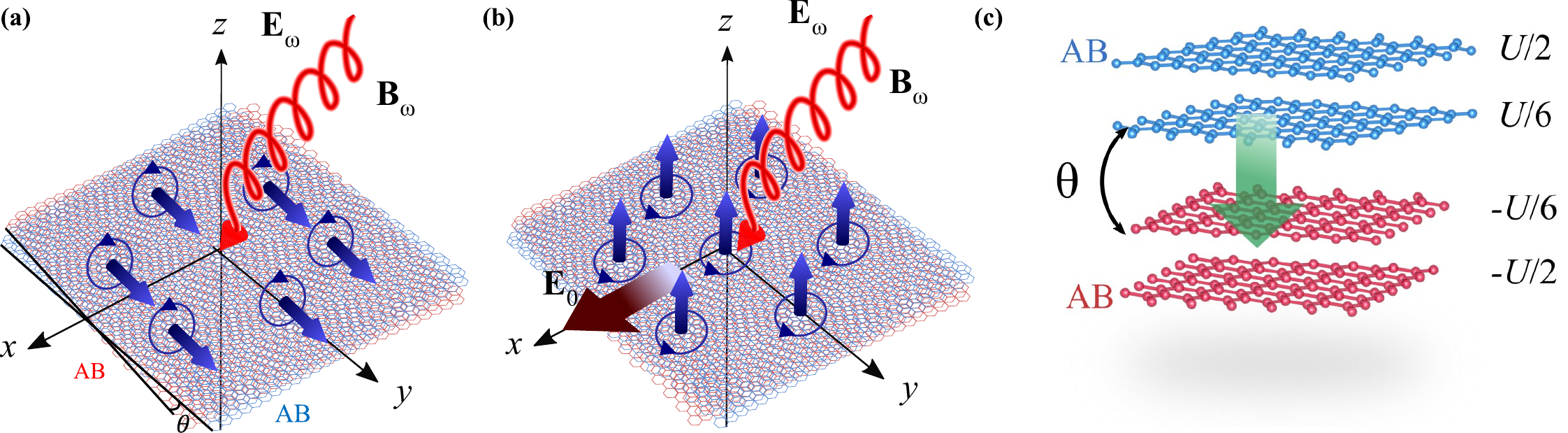}
\caption{(a) In bias-free twisted graphene systems, circular dichroism (CD) has been attributed to the interaction between in-plane magnetic moments $\mathbf{M}$ and the optical fields $\mathbf{E}_\omega$, $\mathbf{B}_\omega$. These in-plane moments arise from chiral interlayer moir\'e coupling at finite twist angles $\theta$.
(b) In contrast, the metallic magnetoelectric electro-optic effect generates out-of-plane magnetic moments in twisted graphene under static electric fields $\mathbf{E}_0$, which couple with the optical fields to produce a distinct dichroic response.
(c) Schematic setup for modeling the vertical displacement field in twisted double bilayer graphene. Following Ref.~\cite{Slot2023}, we neglect screening effects and assume a vertical bias difference of $U/3$ between adjacent graphene layers.}
\label{fig1}
\end{figure*}

In this work, we identify TDBG as an ideal platform for realizing giant, gate-tunable, magnetoelectric EO effects in the terahertz range. We show that the presence of $C_{3z}$ symmetry in TDBG enables leading-order linear magnetoelectric EO response, with moir\'e Bloch states enabling bias-induced magnetoelectric coefficients exceeding $20000$~$\mu_B$/V$\cdot$nm, with a strong dependence on twist angle and vertical displacement field. The resulting circular dichroism (CD) exhibits a distinct angular dependence compared to previous studies in twisted systems~\cite{talkington2023, Kim2016, PhysRevLett.120.046801}, vanishing at normal incidence. While CD in unbiased moiré systems has been attributed to in-plane magnetic moments, as illustrated in Fig.~\ref{fig1}(a), the linear magnetoelectric EO response explored here generates net out-of-plane moments [Fig.~\ref{fig1}(b)], leading to qualitatively different CD signatures. These findings position TDBG as a highly tunable platform for probing Fermi-surface orbital magnetization via optical means and provide a concrete material realization of metallic EO control beyond the conventional Berry curvature dipole (BCD) paradigm.

%%%%%%%%%%%%%%%%%%%%%%%%%%%%%%%%%%%%%%%%%%%%%%%%%%%%%%%%%%%%%%%%%%%%%%%%%%
%%%%%%%%%%%%%%%%%%%%%%%%%%%%%%%%%%%%%%%%%%%%%%%%%%%%%%%%%%%%%%%%%%%%%%%%%%

%%%%%%%%%%%%%%%%%%%%%%%%%%%%%%%%%%%%%%%%%%%%%%%%%%%%%%%%%%%%%%%%%%%%%%%%%%
\begin{figure}[t]
\includegraphics[width = \linewidth]{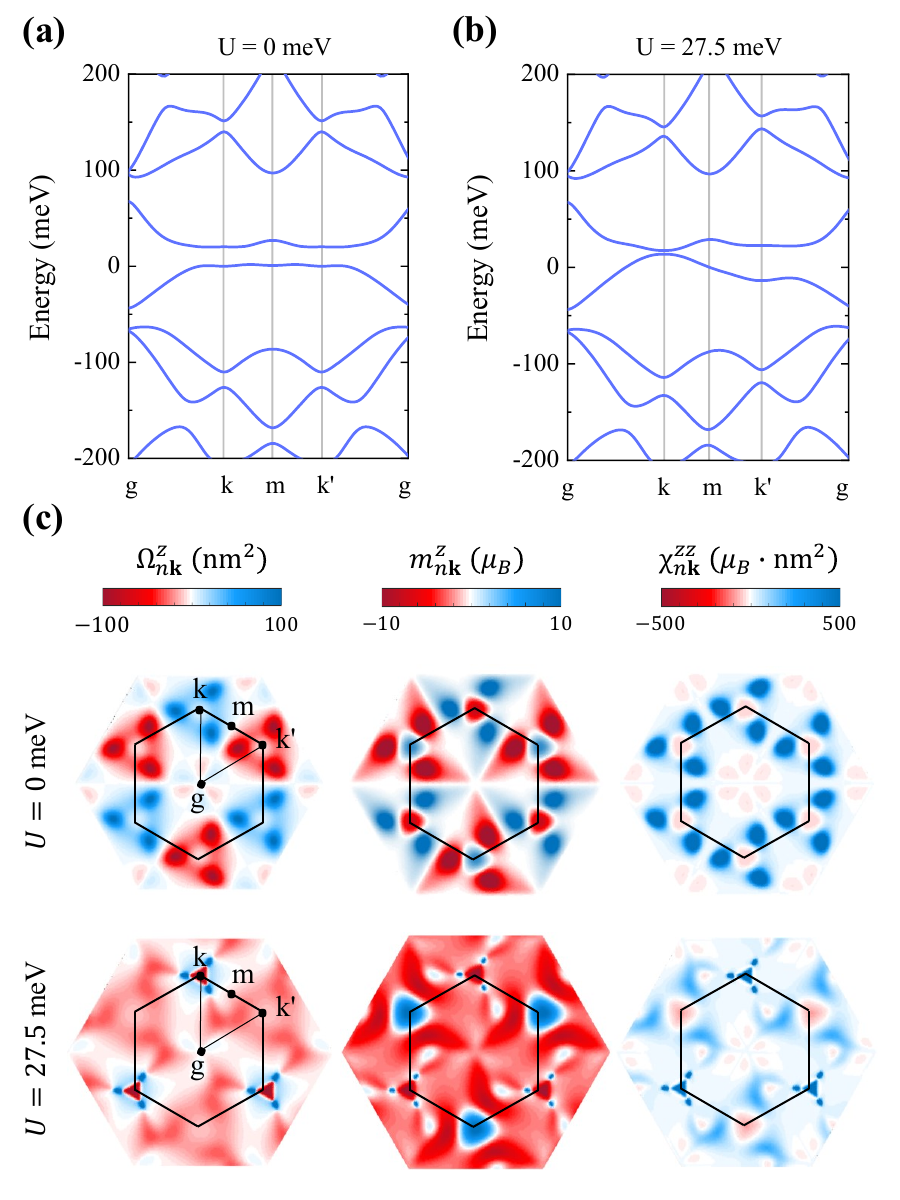}
\caption{(a) and (b) display the electronic structure of double-bilayer graphene twisted by $\theta = 1.75^{\circ}$, under vertical bias $U = 0$ meV and $U = 27.5$ meV, respectively, for the K valley Bistritzer-MacDonald model. (c) Shows the associated momentum-resolved Berry curvature, orbital magnetic moment and the bias-induced magnetoelectric coupling of Bloch electrons of the top most valance states. Results for the K' valley Bistritzer-MacDonald model are obtained by means of a time-reversal operation, i.e., by means of the prescription $\Omega^z_{n\textbf{k}} (\textrm{K-valley}) \rightarrow -\Omega^z_{n\textbf{k}} (\textrm{K'-valley})$, $m^z_{n\textbf{k}} (\textrm{K-valley}) \rightarrow -m^z_{n\textbf{k}} (\textrm{K'-valley})$ and $\chi^{zz}_{n\textbf{k}} (\textrm{K-valley}) \rightarrow \chi^{zz}_{n\textbf{k}} (\textrm{K'-valley})$, where, $\chi^{zz}_{n\textbf{k}} = \Omega^z_{n\textbf{k}}m^z_{n\textbf{k}} $.}
\label{fig2}
\end{figure}
%%%%%%%%%%%%%%%%%%%%%%%%%%%%%%%%%%%%%%%%%%%%%%%%%%%%%%%%%%%%%%%%%%%%%%%%%%

%%%%%%%%%%%%%%%%%%%%%%%%%%%%%%%%%%%%%%%%%%%%%%%%%%%%%%%%%%%%%%%%%%%%%%%%%%
\begin{figure}[t]
\includegraphics[width = \linewidth]{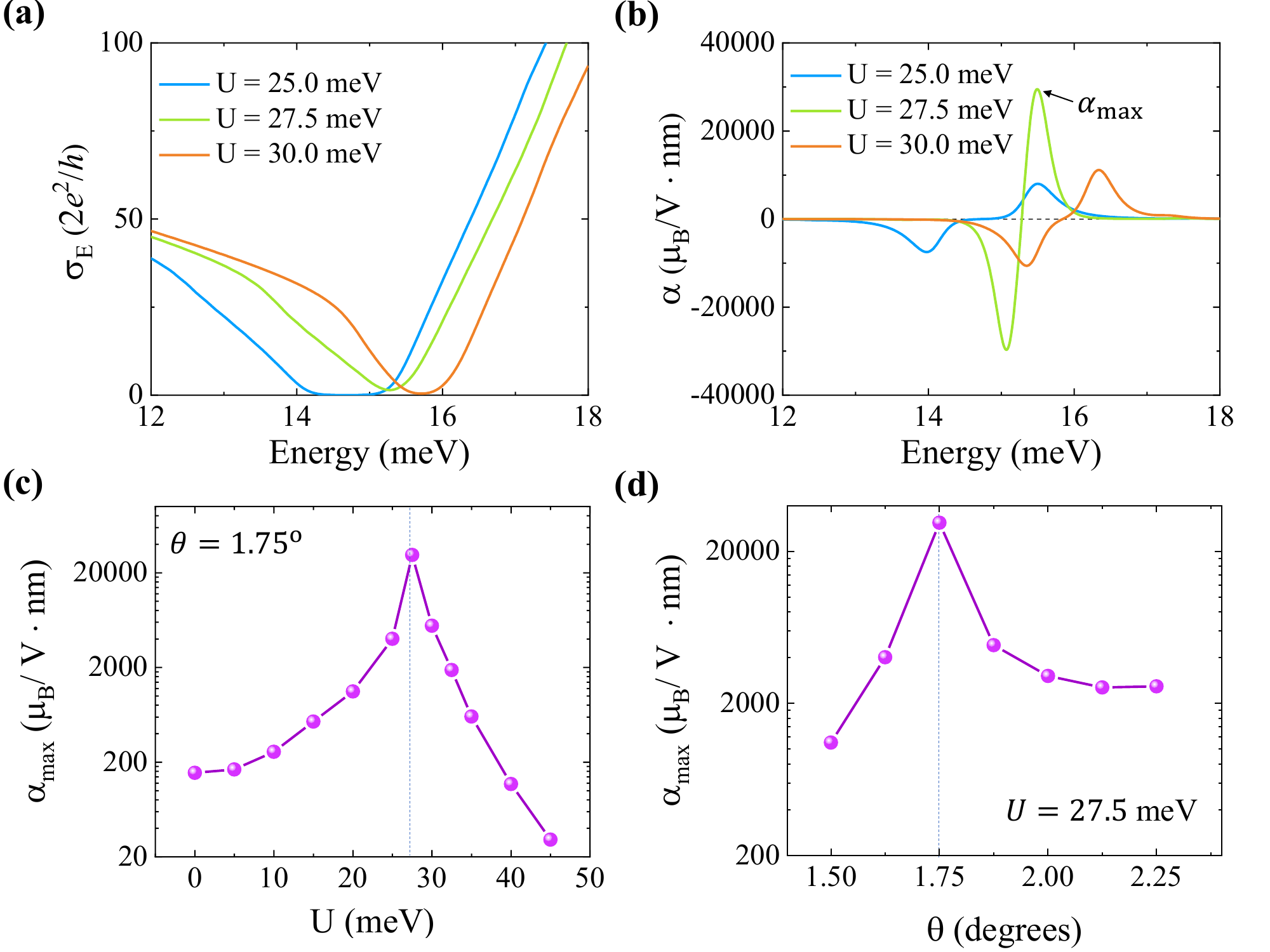}
\caption{The general metallic optical response of $C_{3z}$ symmetric twisted double bilayer graphene, given in Eq.~\ref{eq:constitutive}, depends on two parameters, $\sigma_E$ and $\alpha$, capturing the Drude conductivity and the bias induced metallic magnetoelectric EO response. Panels (a) and (b) display the $\sigma_E = e^2V$ and $\alpha$ as a function of the Fermi energy at distinct twist angles, as obtained from the Bistritzer-MaDonald model. We have set, $\tau = 10$ ps and $E_0^y = 10^4$ V/m. Panels (c) and (d) show the evolution of the maximum attainable bias-induced magnetoelectric coupdling $\alpha_{\textrm{max}}$ as a fuction of the vertical bias $U$ and twist angle $\theta$, respectively. }
\label{fig3}
\end{figure}
%%%%%%%%%%%%%%%%%%%%%%%%%%%%%%%%%%%%%%%%%%%%%%%%%%%%%%%%%%%%%%%%%%%%%%%%%%

\textbf{Metallic EO Responses.} In the low-frequency limit, $\hbar \omega \ll \epsilon_{\textrm{gap}}$, with $\epsilon_{\textrm{gap}}$ being the optical gap, the system's electromagnetic response is dominated by intraband transitions. Within the dilute impurity limit, $\gamma \ll \omega$, where $\tau = 1/\gamma$ is the relaxation time, the semiclassical approach adopted here confines the frequency range to $\gamma \ll \omega \ll \epsilon_{\textrm{gap}}/\hbar$. 

The constitutive relation for inversion-broken and time-reversal symmetric systems takes the form 
\begin{eqnarray}
J_0^{\beta}(\omega) = \sigma^{\alpha\beta}_{\textrm{Drude}}(\omega) E^{\beta}_{\omega} + \sigma^{\alpha\beta}_{\textrm{GME}}(\omega) B^{\beta}_{\omega},
    \label{eq1}
\end{eqnarray}
capturing the conventional AC Drude response and the gyrotropic magnetic effect (GME) derived from the magnetic moment texture of Bloch electrons on the Fermi surface~\cite{PhysRevLett.116.077201}. 

The presence of a static electric field, $\textbf{E}_0$, alters the system’s electromagnetic response by introducing corrections to the constitutive relation, leading to electro-optical (EO) effects~\cite{PhysRevB.110.115421}: $J_0^{\beta}(\omega) \rightarrow J_0^{\beta}(\omega) + J_{\textrm{EO}}^{\beta}$, where $J_{\textrm{EO}}^{\beta}(\omega)$ can be written most generally as 
\begin{eqnarray}
J_{\textrm{EO}}^{\alpha}(\omega) = \sigma^{\alpha\beta}_{\textrm{D}}(\omega) E^{\beta}_{\omega} + \sigma^{\alpha\beta}_{\textrm{G}}(\omega) B^{\beta}_{\omega}.
    \label{eq2}
\end{eqnarray}
Together, the tensors $\boldsymbol{\sigma}_{\textrm{Drude}}(\omega)$, $\boldsymbol{\sigma}_{\textrm{GME}}(\omega)$, $\boldsymbol{\sigma}_{\textrm{D}}(\omega)$ and $\boldsymbol{\sigma}_{\textrm{G}}(\omega)$, account for the relevant optical responses arising from the coupling between optical fields, static bias and the wave function of Bloch electrons in inversion-broken, time-reversal symmetric systems. The nature of these responses has been addressed in previous works~\cite{PhysRevLett.130.076901, PhysRevB.109.245126, PhysRevB.110.115421}. They are summarized as 
\begin{eqnarray}
\boldsymbol{\sigma}_{\textrm{Drude}}(\omega)  = \displaystyle \frac{e^2}{\gamma - i\omega}\textbf{V}, \ \ \ \  \boldsymbol{\sigma}_{\textrm{GME}}(\omega)  = \displaystyle e\frac{i\omega}{i\omega - \gamma}\textbf{K},
    \label{eq3a}
\end{eqnarray}
with $V^{\alpha\beta} = \sum_{n\textbf{k}}(-\partial f^0_{n\textbf{k}}/\partial \epsilon_{n\textbf{k}})v^{\alpha}_{n\textbf{k}}v^{\beta}_{n\textbf{k}}$ and $K^{\alpha\beta} = \sum_{n\textbf{k}}(-\partial f^0_{n\textbf{k}}/\partial \epsilon_{n\textbf{k}})v^{\alpha}_{n\textbf{k}}m^{\beta}_{n\textbf{k}}$, describing unbiased intraband optical responses, 
\begin{eqnarray}
\boldsymbol{\sigma}_{\textrm{D}}(\omega)  = \displaystyle \frac{e^3}{\hbar}\left(\frac{1}{\gamma}\textbf{D} \cdot \textbf{E}_0-\frac{1}{\gamma - i\omega}\textbf{F}_0 \cdot \textbf{D}\right),
\label{eq3b}
\end{eqnarray}
with $D^{\alpha\beta} = \sum_{n\textbf{k}}(-\partial f^0_{n\textbf{k}}/\partial \epsilon_{n\textbf{k}})\Omega^{\alpha}_{n\textbf{k}}v^{\beta}_{n\textbf{k}}$, capturing the contributions derived from the BCD (up to a $\hbar$ factor)~\cite{PhysRevLett.133.146605}, and
\begin{eqnarray}
\boldsymbol{\sigma}_{\textrm{G}}(\omega)  = \displaystyle \frac{e^2}{\hbar} \frac{i\omega}{i\omega - \gamma}\textbf{F}_0 \cdot \textbf{G},
    \label{eq3b}
\end{eqnarray}
with $G^{\alpha\beta} = \sum_{n\textbf{k}}(-\partial f^0_{n\textbf{k}}/\partial \epsilon_{n\textbf{k}})\Omega^{\alpha}_{n\textbf{k}}m^{\beta}_{n\textbf{k}}$, describing the bias-induced correction to the GME, i.e., magnetoelectric EO effect~\cite{PhysRevB.110.115421}. In this work, we will also refer to the quantity $\chi_{n\textbf{k}}^{\alpha\beta} = \Omega^{\alpha}_{n\textbf{k}}m^{\beta}_{n\textbf{k}}$ to quantify the local magnitude of this effect within the Brillouin zone. Here, $\textbf{v}_{n\textbf{k}} = (1/\hbar)\nabla_{\textbf{k}} \epsilon_{n\textbf{k}}$ is the Bloch velocity, $\boldsymbol{\Omega}_{n\textbf{k}}$ is the Berry curvature and $\textbf{m}_{n\textbf{k}}$ is the magnetic moment arising from spin and the self-rotation of Bloch wave packets~\cite{RevModPhys.82.1959}. We have also introduced the fully antisymmetric electric field tensor $\textbf{F}_0$, with components $F_0^{\alpha\beta} = -\epsilon_{\alpha\beta\gamma}E_0^{\gamma}$. 

Unlike previous works that focused solely on the contribution from $\sigma^{\alpha\beta}_{\textrm{D}}(\omega)$~\cite{PhysRevLett.130.076901, PhysRevApplied.22.L041003, PhysRevB.109.245126}, this study underscores the critical role of the bias-induced magnetoelectric term, $\sigma^{\alpha\beta}_{\textrm{G}}(\omega)$, offering a more complete picture of metallic EO responses. In this work, we investigate TDBG~\cite{Haddadi2020, PhysRevB.99.235406, Shen2020, Slot2023, PhysRevB.99.235417, PhysRevB.100.201402} as a promising platform for metallic EO effects. Remarkably, we demonstrate that these systems exhibit \textit{leading-order} giant magnetoelectric EO responses due to their point group symmetries, which enforce $\textbf{D} = \textbf{0}$ and $\textbf{K} = \textbf{0}$. In the following, we discuss the symmetry-enforced form of the metallic EO response tensors.

\textbf{Symmetry-enforced metallic EO responses.} Before addressing the electronic structure and the optical responses induced by a static electric field in this system, we first focus on the symmetry-dictated form of the response tensors. The moir\'e pattern of TDBG, shown in Figs.~\ref{fig1}(a) and (b), display $C_{3z}$ symmetry. Thus, we have studied the impact of $C_{3z}$ symmetry on the Berry curvature dipole components $D^{zx}$ and $D^{zy}$. However, note that the following analysis applies equally to the corresponding components of the gyrotropic magnetic tensor, $\textbf{K}$, due to its symmetry equivalence to $\textbf{D}$. 

The point group of twisted (double) bilayer graphene, which imposes constraints on the components of the magnetoelectric response tensors, depends on the type of stacking and twist angle~\cite{PhysRevB.98.085435, SI}. The point group associated with the TDBG system considered here is $D_3$~\cite{Koshino_TDBG}, which is generated by a threefold axis along the stacking direction, $C_{3z}$, and an in-plane twofold axis, $C_{2y}$. As a consequence of the oddness  of $\Omega ^z_{n\textbf{k}}$ and evenness of $v^y_{n\textbf{k}}$ under $C_{2y}$, their product is odd. Noting that the first Brillouin zone (BZ) and the energy derivative of the Fermi function are invariant with respect to the point group symmetries of the crystal, integration over the BZ indicates $D^{zy} = 0$~\cite{SI}. However, the oddness of $v ^x_{n\textbf{k}}$ under this symmetry indicates that $D^{zx} $ is symmetry-allowed  by $C_{2y}$. We will now show that the presence of a $C_{3z}$ symmetry alone is sufficient to enforce the vanishing of $D^{zx}$ and $D^{zy}$.
 
 $C_{3z}$ is a proper rotation, so the Berry curvature and velocity components transform in an identical manner. Accounting for this symmetry and integrating over the BZ gives $D^{zx}_{n} = -\frac{\sqrt{3}}{3}D^{zy}_n$ and $D^{zy}_{n} = \frac{\sqrt{3}}{3}D^{zx}_n$~\cite{SI}. Thus, $D^{zx}_{n} = D^{zy}_{n} = 0$. We have systematically performed a symmetry analysis on all graphene systems of interest beginning with the monolayer case and building up to TDBG in levels of decreasing symmetry (please see Supplementary Materials~\cite{SI}). Our analysis shows that all cases must have vanishing $D^{zx}_n$ and $D^{zy}_n$ by virtue of $C_{3z}$. 

As we have shown, nonzero $D^{z\beta}$, $\beta = x,y$, and the corresponding linear magnetoelectric coupling components require $C_{3z}$ symmetry breaking, commonly achieved through strain~\cite{Sinha2022, PhysRevLett.123.036806}. Hence, in the absence of strain, the non-linear magnetoelectric electro-optical effect discussed here is the leading order magnetoelectric response. It is important to note that, linear GME ($\propto E_{\omega}^{\alpha}$) is not possible within the TDBG point group, but the non-linear ($\propto E_0^{\gamma}E_{\omega}^{\alpha}$) counterpart is. In the following, we study the electronic structure properties enabling sizable metallic magnetoelectric EO effects in this system.

\textbf{Metallic EO responses of TDBG.} In this section, we focus on the electronic structure of TDBG, highlighting quantities relevant to the magnetoelectric EO effect. We adopt the Bistritzer-MacDonald continuum (BM) model to address the electronic structure of TDBG~\cite{Bistritzer2011, Slot2023, Snote}. The electrostatic potential due to a vertical displacement field is modeled as a linear potential drop of magnitude $U$, as shown in Fig.~\ref{fig1}(c), where screening effects have been neglected. For simplicity, we explicitly focus on the $K$-valley BM model, noting that the corresponding results for the $K'$-valley can be obtained via time-reversal symmetry. We confine our description to small deviations off the $1.75^{\circ}$ configuration, to ensure that our single-particle model is an accurate description~\cite{Slot2023}. The electronic band structure of the $1.75^{\circ}$ TDBG at a vanishing displacement field ($U = 0$ meV) is shown in Fig.~\ref{fig2}(a). The moir\'e interlayer coupling leads to narrow low-energy bands, with band widths of $\approx 50$~meV for the top valence and bottom conduction states, separated by an energy gap of $\approx 20$ meV.

A finite vertical displacement field radically alters the electronic structure of TDBG~\cite{Koshino_TDBG, PhysRevB.99.235417, PhysRevB.100.201402}. As the magnitude of the bias $|U|$ increases, it drives a gap closure followed by a reopening at either the k or k' valley of the moir\'e Brillouin zone, determined by the sign of $U$~\cite{Snote}. Figure~\ref{fig2}(b) displays the band structure in the vicinity of the gap-closing regime for $U = 27.5$ meV, where the gap at the k point is reduced to $\approx3.7$ meV, while it is enhanced to $\approx33$ meV at the k' point. Further increase of $U$ induces a gap reopening at the k point.

This high degree of tunability in the electronic structure via $U$ has a direct influence on both the Berry curvature and the orbital magnetic moment of the Bloch states. Figure~\ref{fig2}(c) presents the momentum-resolved maps of the out-of-plane components of the Berry curvature $\Omega^z_{n\textbf{k}}$, orbital magnetic moment $m^z_{n\textbf{k}}$, and their product $\chi^{zz}_{n\textbf{k}} = \Omega^z_{n\textbf{k}}m^z_{n\textbf{k}}$ (the magnetoelectric EO effect integrand), evaluated at mid-gap for $U = 0$ meV (top row) and $U = 27.5$ meV (bottom row). At zero bias, $\Omega^z_{n\textbf{k}}$ and $m^z_{n\textbf{k}}$ exhibit symmetric distributions in magnitude, with opposite signs centered around the k and k' valleys. The presence of a finite displacement field ($U = 27.5$ meV), however, induces a marked asymmetry: both quantities become strongly localized in momentum space around the k valley, with a corresponding suppression near the k' valley. This redistribution reflects the bias-induced tendency for gap closure at k and enhancement at k', as observed in the band structure. The same features are also observed for $\chi^{zz}_{n\textbf{k}}$. 

Recall that the results shown in Fig.~\ref{fig2}(c) correspond to the $K$-valley of each constituent bilayer graphene. The analogous quantities for the $K'$-valley are related by time-reversal symmetry, which imposes $\boldsymbol{\Omega}^{K}_{n\textbf{k}} = -\boldsymbol{\Omega}^{K'}_{n\textbf{k}}$ and $\textbf{m}^{K}_{n\textbf{k}} = -\textbf{m}^{K'}_{n\textbf{k}}$. While both the Berry curvature and the orbital magnetic moment are odd under time-reversal, their product—the magnetoelectric response tensor $\boldsymbol{\chi}_{n\textbf{k}} = \boldsymbol{\Omega}_{n\textbf{k}}  \textbf{m}_{n\textbf{k}}$—is even. As a result, time-reversal symmetry prohibits the emergence of a net anomalous Hall current or total orbital magnetization, as expected, but permits non-vanishing magnetoelectric EO responses.

%%%%%%%%%%%%%%%%%%%%%%%%%%%%%%%%%%%%%%%%%%%%%%%%%%%%%%%%%%%%%%%%%%%%%%%%%%%%%%
\begin{figure}
    \includegraphics[width=1\linewidth]{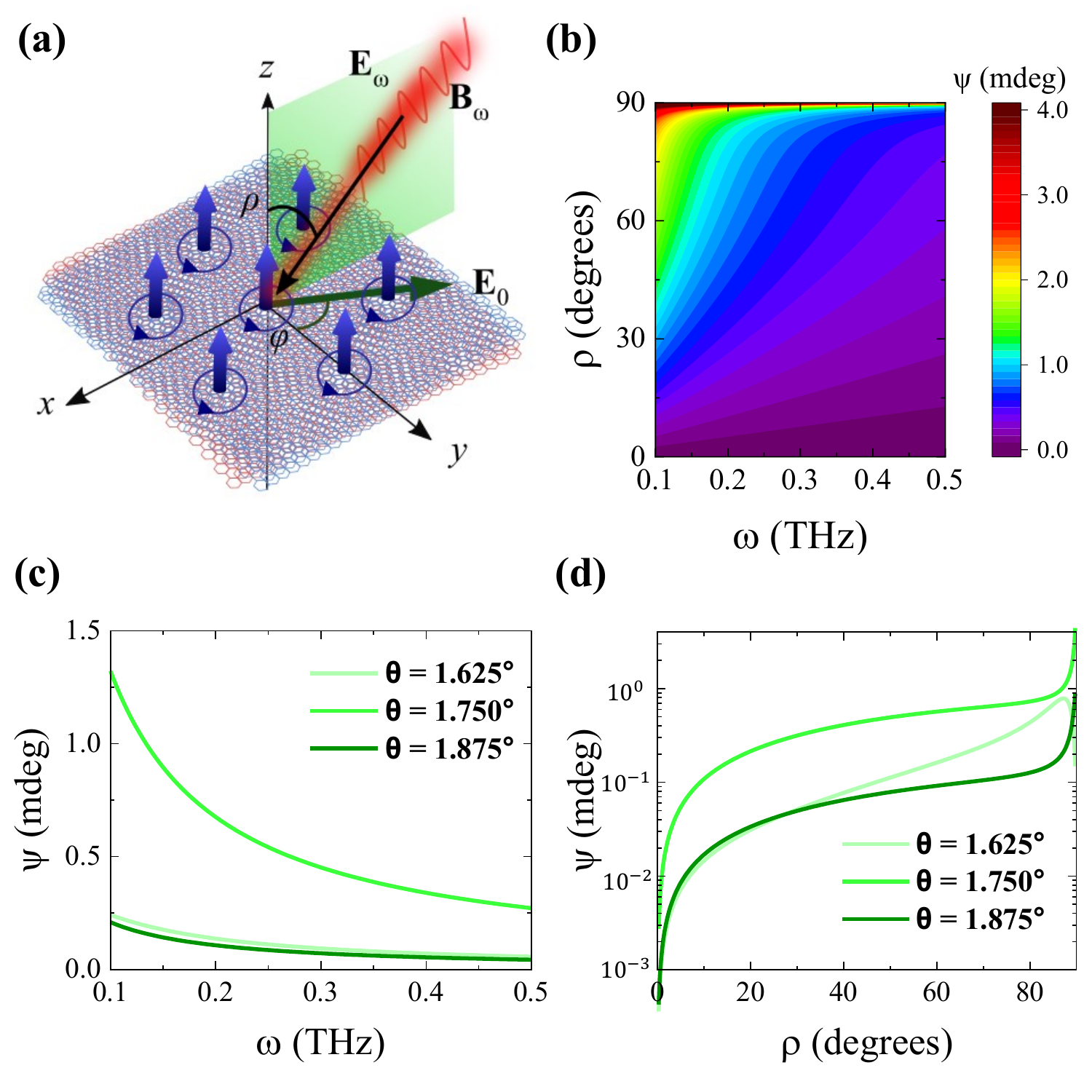}
    \caption{(a) Schematic of the system with the twisted double bilayer graphene at $z=0$, in between dielectrics with relative permittivities $\epsilon_1$ ($z < 0$) and $\epsilon_2$ ($z > 0$). The incidence plane is depicted and the incidence angle is $\rho$ (b) Ellipticity with respect to frequency and incidence angle $\rho$ at $\theta = 1.750^{\circ}$ twisting; (c) with respect to frequency at $\rho = 45^{\circ}$; and (d) with respect to $\rho$ at $\omega = 0.3$ THz (See Fig.~\ref{fig3} for corresponding $\sigma_\textrm{E}$ and $\alpha$). In panels (b)-(d), $\gamma = 10^{11}$ rad$\cdot$s$^{-1}$, $\epsilon_1 = \epsilon_2 = 1$, $E_0 = 10^5$ V$\cdot$m$^{-1}$, $\phi = 90^{\circ}$.}
    \label{fig:plots}
\end{figure}
%%%%%%%%%%%%%%%%%%%%%%%%%%%%%%%%%%%%%%%%%%%%%%%%%%%%%%%%%%%%%%%%%%%%%%%%%%%%%%

Given these salient features, we discuss now their impact on the intraband optical conductivity of TDBG in the presence of $\textbf{E}_0$. The total metallic response of the system is dictated by the Drude conductivity $\boldsymbol{\sigma}_{\textrm{Drude}}(\omega)$ and the magnetoelectric EO conductivity $\boldsymbol{\sigma}_{\textrm{G}}(\omega)$. To quantify the Drude response, we define the frequency-independent quantity $\sigma_E = e^2V$, where $V$ is a diagonal element of the $\textbf{V}$ tensor (because the system is isotropic $V^{xx}=V^{yy}=V$). Figure~\ref{fig3}(a) shows $\sigma_E$ obtained from the BM model for a few values of $U$ near 27.5 meV, assuming $\gamma = 10^{-11}$ rad/s ($\tau = 10$ ps). In addition to magnitude, these results qualitatively reflect the behavior across the gap-closing and reopening regime, which will become important in upcoming analysis. To quantify the magnetoelectric EO response, we define the frequency-independent quantity $\alpha = (e^2/\gamma\hbar)E_0 G^{zz}$, where $G^{zz} = \sum_{n\textbf{k}}(-\partial f^0_{n\textbf{k}}/\partial \epsilon_{n\textbf{k}}) \chi^{zz}_{n\textbf{k}}$, is the $zz$ components of the magnetoelectric EO tensor. The quantity $\alpha$ is the effective magnetoelectric coupling coefficient induced by $E_0 = |\textbf{E}_0|$. Figure~\ref{fig3}(b) presents the results for $\alpha$ obtained from the BM model for the same set of values of $U$ near 27.5 meV. Following Ref.~\cite{He2020}, we assume $E_0 = 10^4$ V/m. The results show that $\alpha$ peaks near the energy gap and changes sign between electron and hole bands. 

The behavior of the BM model can be captured by an effective model describing states near the band gap at the k point, where a gapped Dirac Hamiltonian offers an accurate representation for the band structure~\cite{Snote}. Within this framework, we obtain $G^{zz} = e (\hbar v_F \Delta)^2 / 32\pi\hbar \mu^4$ for the conduction band ($\mu > \Delta/2$) and $G^{zz} = -e (\hbar v_F \Delta)^2 / 32\pi\hbar \mu^4$ for the valence band ($\mu < -\Delta/2$), where $\Delta$ is the bandgap, $v_F$ is the Fermi velocity, and $\mu$ is the chemical potential. The pronounced $1/\mu^4$ dependence accounts for the sharp peaks observed near the band edges. The model predicts that the maximum possible value occurs precisely at $\mu = \pm\Delta/2$, with a magnitude scaling as $|G^{zz}| \propto (\hbar v_F / \Delta)^2$. Therefore, a smaller gap leads to a larger $\alpha$. This trend is clearly reflected in Fig.~\ref{fig3}(b), where the magnitude of $\alpha$ peaks at $U = 27.5$ meV, corresponding to the configuration nearest to gap closure. Note that the spectral broadening present in the numerical simulations displaces the peak in $\alpha$ slightly away from the band edge, resulting in the smoothened profile seen in the figure.

Remarkably, the maximum attainable bias-induced magnetoelectric coefficient, $\alpha_{\textrm{max}}$, reaches exceptionally large values as high as $20000$~$\mu_B$/V$\cdot$nm, where $\mu_B$ denotes the Bohr magneton, under a moderate external field of $E_0 = 10^4$ V/m. This value exceeds, by roughly an order of magnitude, the giant magnetoelectric response associated with the $\mathbf{K}$ tensor in strained twisted bilayer graphene, as reported in Ref.~\cite{He2020}. The highly tunable electronic structure of TDBG translates into a remarkable degree of control over the magnetoelectric response. To illustrate this, the dependence of $\alpha_{\textrm{max}}$ on the vertical displacement field $U$ and twist angle $\theta$ is shown in Fig.~\ref{fig3}(c) and (d). While $\alpha_{\textrm{max}}$ is strongly sensitive to $U$, it saturates around $2000$~$\mu_B$/V$\cdot$nm for twist angles $\theta > 2^\circ$ at $U = 27.5$ meV. These results highlight TDBG as an exceptional and highly tunable platform for realizing metallic magnetoelectric EO effects. In the next section, we provide a practical analysis of how this effect manifests in optical dichroism, outlining a clear experimental route for its detection.

\textbf{Bias-Induced Circular Dichroism in TDBG}. In this section, we examine how the giant metallic magnetoelectric EO effect gives rise to circular dichroism. The configuration is illustrated in Fig~\ref{fig:plots}(a), where $\rho$ denotes the angle of incidence and $\phi$ defines the orientation of the in-plane static electric field $\textbf{E}_0$ with respect to the $xz$-plane of incidence. While our analysis focuses on wave propagation within the $xz$-plane, the general scenario can be recovered by varying the direction of the in-plane bias $\textbf{E}_0$. The TDBG is positioned at $z = 0$, sandwiched between two dielectric media with equal permittivity $\epsilon_1 = \epsilon_2 = \epsilon = 1$. The optical fields $\textbf{E}_\omega$ and $\textbf{B}_\omega$ propagate along the $+z$-direction. Our approach follows the scattering formalism introduced in Ref.~\cite{oliva2019}, with a detailed derivation provided in the Supplemental Material~\cite{Snote}. The total effective conductivity, $\boldsymbol{\sigma}_{\textrm{eff}}(\omega)$, in the $\textbf{E}_\omega = E_\omega^x \hat{\textbf{x}} + E_\omega^y \hat{\textbf{y}}$ basis is
\begin{equation}
  \boldsymbol{\sigma}_{\textrm{eff}} = \frac{\sigma_E}{\gamma - i\omega} \begin{bmatrix}
        1 & 0 \\
        0 & 1
    \end{bmatrix} + \frac{i\omega \alpha}{\gamma-i\omega}\left(\frac{\gamma\epsilon}{c}\right)\sin\rho\begin{bmatrix}
        0 & \sin\phi \\
        0 & -\cos{\phi}
    \end{bmatrix},
    \label{eq:constitutive}
\end{equation}
where $c$ denotes the speed of light, and $\gamma = 10^{11}$~rad$\cdot$s$^{-1} \approx 0.02~\text{THz}$ represents the adopted scattering rate. We restrict the frequency of the incident wave to the range $0.02~\textrm{THz} \ll \omega \ll 0.9~\textrm{THz}$—with the lower bound ensuring validity within the dilute impurity limit, and the upper bound avoiding the onset of the interband transition regime. The latter is defined by the smallest optical gap $\epsilon_{\textrm{gap}} = 3.7$ meV $\approx 0.9$ THz, corresponding to a twist angle $\theta = 1.750^\circ$ in TDBG under a vertical bias of $U = 27.5$ meV. The effective conductivity parameters, $\sigma_E$ and $\alpha$, were obtained from the BM model as outlined in the previous section. Following Refs.~\cite{andrews2020, rodger1997} we define the ellipticity as $\psi \approx 32.982\cdot (A_L - A_R)$, where $A_{L}$ and $A_{R}$ denotes the absorbance of left-circularly polarized (LCP) and right-circularly polarized (RCP) light, respectively, and the prefactor of 32.982 converts the final result to degrees. We present our results in millidegrees, following the conventions of Refs.~\cite{talkington2023, Kim2016, addison2019}. 

The circular dichroism is caused by the imaginary component of the off-diagonal conductivity, i.e. $\text{Im}(\sigma^{xy}_{\textrm{eff}})$ in Eq.~(\ref{eq:constitutive}), which induces a current in the $+x$-direction ($-x$) in response to RCP (LCP) light. In this work, we assume $\phi = 90^\circ$ in order to maximize $\sigma^{xy}_{\textrm{eff}}$. Figure~\ref{fig:plots}(b) shows that the dichroism decreases with frequency but increases with incidence angle. The former case can be understood by noting that $\text{Im}(\sigma^{xy}_{\textrm{eff}}) \rightarrow 0$ in the large frequency limit $\omega \rightarrow \infty$. This behavior is more clearly illustrated in Fig.~\ref{fig:plots}(c), which also compares the results for TDBG twisted at $\theta = 1.625^{\circ}$, $\theta = 1.750^{\circ}$, and $\theta = 1.875^{\circ}$, all evaluated at a fixed incidence angle of $\rho = 45^\circ$. Notably, the ellipticities obtained here are comparable to the THz circular dichroism observed in twisted bilayer graphene in Ref.~\cite{talkington2023}, as well as to the pronounced dichroism reported in the visible to near-ultraviolet range in multilayer twisted graphene systems~\cite{Kim2016}, both measured at normal incidence. In contrast to our study, these effects are attributed to in-plane magnetic moments~\cite{Kim2016} and do not originate from a static field $\textbf{E}_0$.

Furthermore, the coupling with the $B^z_{\omega}$ component of the optical field through $m^z_{n\textbf{k}}$ gives rise to two competing effects with respect to $\omega$: a Drude-like contribution to the non-equilibrium distribution function of electrons, proportional to $1/(\gamma - i\omega)$ capturing the balance between relaxation and the optical field tendency to drive the system out-of-equilibrium, and a Zeeman-like ($\propto B^z_{\omega}m^z_{n\textbf{k}} e^{i\omega t}$) correction to the Bloch energy $\epsilon_{n\textbf{k}}$, proportional to $i\omega$~\cite{PhysRevLett.116.077201,PhysRevB.110.115421}. The Drude-like response dominates in the low-frequency regimes, whereas the Zeeman-like response dominates as the frequency increases. At lower frequencies, $\text{Im}(\sigma^{xy}_{\textrm{eff}})$ is larger (stronger dichroic response), whereas at higher frequencies, $\text{Im}(\sigma^{xy}_{\textrm{eff}})$ is lower (weaker dichroic response). 

The enhancement of circular dichroism with increasing incidence angle is explicitly shown in Fig.~\ref{fig:plots}(d), and is now discussed in greater detail. This behavior originates from the Zeeman coupling between $B^z_\omega$ and $m^z_{n\textbf{k}}$, which increases with $\rho$ in our setup, as can be seen in Eq.~(\ref{eq:constitutive}). Note that this coupling causes the circular dichroism to vanish at normal incidence, since $B_\omega^z = 0$ when $\rho = 0$. This feature offers a unique signature of optical dichroism induced by magnetoelectric EO effects, in striking contrast to previous studies~\cite{talkington2023, Kim2016}. We also find that at large angles of incidence, the magnitude of the Drude conductivity plays a role in determining the circular dichroism spectra. For TDBG at $\theta = 1.625^{\circ}$ a broader and smaller peak is observed in the spectra, which is due to the large Drude conductivity compared to the $\theta = 1.750^{\circ}$ and $\theta = 1.875^{\circ}$ cases (See Fig.~\ref{fig3}). The higher dissipative Drude component decreases the peak and the higher reactive component broadens it~\cite{Snote}.

\textit{Acknowledgments.} 
D. S., C.O.A., and T. L. acknowledge support from Office of Naval Research MURI grant N00014-23-1-2567. N. RL acknowledges support form the University of Minnesota Pathways to Graduate School: Summer Research Program.

\bibliographystyle{apsrev}% your bst file here
\bibliography{my.bib} %your bib file here

\end{document}